\documentclass[12pt]{article}
\usepackage[dvips]{graphicx}

\usepackage{graphicx}
\newcommand{\text}{\rm}

\newcommand{\ug}{ \; = \; }

\newcommand{\bb}{\begin{equation}}
\newcommand{\ee}{\end{equation}}
\newcommand{\bega}{\begin{eqnarray}}
\newcommand{\ega}{\end{eqnarray}}
\newcommand{\begae}{\begin{eqnarray*}}
\newcommand{\egae}{\end{eqnarray*}}

\newcommand{\h}{\hspace*{4ex}}
\newcommand{\dis}{\displaystyle}

\newcommand{\be}{\beta}

\newcommand{\om}{\omega}

\newcommand{\cent}{\centerline}
\newcommand{\vs}{\vspace*}

\begin{document}

\baselineskip 0.7cm

\begin{center}
{\Large {\bf Frozen Waves: Stationary optical wavefields with arbitrary
longitudinal shape, by superposing equal frequency Bessel beams:
Frozen Waves}$^{\: (\dag)}$} \footnotetext{$^{\: (\dag)}$  Work
supported by FAPESP (Brazil), ; previously available as e-print
******. \ E-mail address for contacts:
mzamboni@dmo.fee.unicamp.br}

\end{center}

\vs{4mm}

\cent{ Michel Zamboni-Rached, }

\vs{0.3 cm}

\centerline{{\em Department of Microwaves and Optics, Faculty of
Electrical Engineering,}} \centerline{{\em State University
of Campinas, Campinas, SP, Brazil.}}

\vs{0.5 cm}

\

{\bf Abstract  \ --} \  In this paper it is shown how one can use
Bessel beams to obtain a {\em stationary} localized wavefield
with high transverse localization, and whose longitudinal
intensity pattern can assume any desired shape within a chosen
interval $0\leq z \leq L$ of the propagation axis. This intensity
envelope remains static, i.e., with velocity $v=0$; and because of
this we call ``Frozen Waves" such news solutions to the wave
equations (and, in particular, to the Maxwell equations).  These
solutions can be used in many different and interesting
applications, as optical tweezers, atom guides, optical or
acoustic bistouries, various
important medical purposes, etc.\\



\

{\bf 1. -- Introduction}

\h Since many years a theory of localized waves (LW), or
nondiffracting waves, has been developed, generalized, and
experimentally verified in many fields as optics, microwaves and
acoustics[1]. These waves have the surprising characteristics of
resisting the diffraction effects for long distances, i.e., of
possessing a large depth of field.

\h These waves can be divided into two classes, the localized
beams, and the localized pulses.  With regard to the beams, the
most popular is the Bessel beam[1].

\h Much work have been made about the properties and applications
of single Bessel beams.  By contrast, only a few papers have been
addressed to the properties and applications of {\em
superpositions} of Bessel beams with the same frequency, but with
different longitudunal wave numbers. The few works on this subject
have shown some surprising possibilities related with this type of
superpositions, mainly the possibility of controlling the
transverse shape of the resulting beam[2,3]. The other important
point, i.e., that of controlling the longitudinal shape, has been
very rarely analyzed, and the relevant papers have been confined
to numerical optimization processes[4,5] to find out one
appropriate computer-generated hologram.

\h In this work we develop a very simple method$^{**}$ that makes
possible the control of the beam intensity longitudinal shape
within a chosen interval $0\leq z \leq L$, where $z$ is the
propagation axis and $L$ can be much greater than the wavelength
$\lambda$ of the monochromatic light which is being used. Inside
such a space interval, we can construct a {\em stationary}
envelope with many different shapes, including one or more
high-intensity peaks (with distances between them much larger than
$\lambda$). This intensity envelope remains static, i.e., with
velocity $V=0$; and because of this we call ``Frozen Waves" such
news solutions to the wave equations (and, in particular, to the
Maxwell equations).

\h We also suggest a simple apparatus capable of generating these
stationary fields.

\h Static wave solutions like these can have many different and
interesting applications, as optical tweezers, atom guides,
optical or acoustic bisturies, electromagnetic or ultrasound
high-intensity fields for various important medical purposes,
etc..$^{**}$ \footnotetext{$^{**}$ Patent pending.}

\

{{\bf 2. -- The mathematical methodology} $^{**}$}

\h We start with the well known axis-symmetric Bessel beam

\bb \psi(\rho,z,t)\ug J_0(k_{\rho}\rho)e^{i\be z}e^{-i\om t}
\label{bb}\ee

with

\bb k_{\rho}^2=\frac{\om^2}{c^2} - \be^2 \; , \label{k}  \ee

where $\om$, $k_{\rho}$ and $\be$ are the angular frequency, the
transverse and the longitudinal wave numbers, respectively. We also
impose the conditions

\bb \om/\be \geq 0 \;\;\; {\rm and}\;\;\; k_{\rho}^2\geq 0
\label{c2} \ee

to ensure forward propagation only, as well as a physical behavior of
the Bessel function.

\h Now, let us make a superposition of $2N + 1$ Bessel beams with
the same frequency $\om_0$, but with {\em different} (and still unknown)
longitudinal wave numbers $\be_n$:

\bb \dis{\Psi(\rho,z,t) \ug e^{-i\,\om_0\,t}\,\sum_{n=-N}^{N}
A_n\,J_0(k_{\rho\,n}\rho)\,e^{i\,\be_n\,z} } \; , \label{soma} \ee

where $A_n$ are constant coefficients.  For
each $n$, the parameters $\om_0$, $k_{\rho\,n}$ and $\beta_n$ must
satisfy Eq.(\ref{k}), and, because of conditions (\ref{c2}), when
considering $\om_0 > 0$, we must have

\bb 0 \leq \be_n \leq \frac{\om_0}{c} \label{be} \ee

\h Now our goal is to find out the values of the longitudinal
wave numbers $\be_n$ and of the coefficients $A_n$ in order to reproduce
approximately, inside the interval $0 \leq z \leq L$ (on the axis $\rho=0$),
a chosen longitudinal intensity pattern that we call $|F(z)|^2$. \
In other words, we want to have

\bb \sum_{n=-N}^{N} A_n e^{i\,\be_n\,z} \approx F(z) \;\;\;\; {\rm
with}\;\;\; 0\leq z \leq L   \label{soma1} \ee

\h Following Eq.(\ref{soma1}), one might be tempted to take $\be_n
= 2\pi n/L$, thus obtaining a truncated Fourier series, expected
to represent the desired pattern $F(z)$. \ Superpositions of
Bessel beams with $\be_n = 2\pi n/L$ has been actually used in
some works to obtain a large set of {\it transverse} amplitude
profiles[2]. However, for our purposes, this choice is not
appropriate due two principal reasons: \ 1) It yields negative
values for $\be_n$ (when $n<0$), which implies backwards
propagating components (since $\om_0 > 0$); \ 2) In the cases when
$L>>\lambda_0$, which are of our interest here, the main terms of
the series would correspond to very small values of $\be_n$, which
results in a very short field depth of the corresponding Bessel
beams(when generated by finite apertures), impeding the creation
of the desired envelopes far form the source.

\h Therefore, we need to make a better choice for the values of
$\be_n$, which allows forward propagation components only, and
a good depth of field. \ This problem can be solved by putting

\bb \be_n \ug Q + \frac{2\,\pi}{L}\,n \label{be2} \; , \ee

where $Q>0$ is a value to be chosen (as we shall see) according to the
given experimental situation, and the desired degree of
transverse field localization. \ Due to Eq.(\ref{be}), we get

\bb 0\leq Q \pm \frac{2\,\pi}{L}\,N \leq \frac{\om_0}{c} \label{N}
\ee

Inequation (\ref{N}) determines the maximum value of $n$, that
we call $N$, once $Q$, $L$ and $\om_0$ have been chosen.

\h As a consequence, for getting a longitudinal intensity pattern
approximately equal to the desired one, $F(z)$, in the interval
$0\leq z \leq L $, Eq.(\ref{soma}) should be rewritten as:

\bb \dis{\Psi(\rho=0,z,t) \ug
e^{-i\,\om_0\,t}\,e^{i\,Q\,z}\,\sum_{n=-N}^{N}
A_n\,e^{i\,\frac{2\pi}{L}n\,z} } \; , \label{soma2} \ee

with

\bb A_n \ug \frac{1}{L} \dis{
\int_{0}^{L}\,F(z)\,e^{-i\,\frac{2\pi}{L}\,n\,z}\,d\,z }
\label{An}\ee

\h Obviously, one obtains only an approximation to the desired
longitudinal pattern, because the trigonometric series
(\ref{soma2}) is necessarily truncated. Its total number of terms,
let us repeat, will be fixed once the values of $Q$, $L$ and $\om_0$
are chosen.

\h When $\rho \neq 0$, the wave field $\Psi(\rho,z,t)$ becomes

\bb \dis{\Psi(\rho,z,t) \ug
e^{-i\,\om_0\,t}\,e^{i\,Q\,z}\,\sum_{n=-N}^{N}
A_n\,J_0(k_{\rho\,n}\,\rho)\,e^{i\,\frac{2\pi}{L}n\,z} } \; ,
\label{soma3} \ee

with

\bb k_{\rho\,n}^2 \ug \om_0^2 - \left(Q + \frac{2\pi\,n}{L}
\right)^2 \ee

\h The coefficients $A_n$ will yield the amplitudes and the
relative phases of each Bessel beam in the superposition.

\h Because we are adding together zero order Bessel functions, we
can expect a {\em high} field concentration around $\rho=0$.

\

{\bf 3. -- Some examples }

\h In this section we shall present two examples of our methodology.

\h Let us suppose that we want an optical wave field with
$\lambda_0 = 0.632 \, \mu$m, that is, with $\om_0 = 2.98 \, 10^{15}\,$Hz),
whose longitudinal pattern (along its $z$-axis) in the range $0 \leq z \leq
L$ is given by the function

 \bb
 F(z) \ug \left\{\begin{array}{clr}
 -4\,\,\dis{\frac{(z-l_1)(z-l_2)}{(l_2 - l_1)^2}} \;\;\; & {\rm for}\;\;\; l_1 \leq z \leq l_2  \\
\\
 \;\;\;\;\;\;\;\;1 & {\rm in}\;\;\; l_3 \leq z \leq l_4 \\
\\
 -4\,\,\dis{\frac{(z-l_5)(z-l_6)}{(l_6 - l_5)^2}} & {\rm for}\;\;\; l_5 \leq z \leq
 l_6 \\
 \\
 \;\;\;\;\;\;\;\; 0  & \mbox{elsewhere} \ ,
\end{array} \right. \label{Fz1}
 \ee

\

where $l_1=L/10$, \ $l_2=3L/10$, \ $l_3=4L/10$, \ $l_4=6L/10$, \
$l_5=7L/10$ and $l_6=9L/10$. In other words, the desired
longitudinal shape, in the range $0 \leq z \leq
L$, is a parabolic function for $l_1 \leq z \leq l_2$, a unitary step
funtion for $l_3 \leq z \leq l_4$, and again a parabola in
the interval $l_5 \leq z \leq
 l_6$, it being zero elsewhere (in the interval $0 \leq z \leq
L$). In this example, let us put $L=0.5\,$m.

\h We can then calculate the coefficients $A_n$, which appear in
the superposition (\ref{soma3}), by inserting Eq.(\ref{Fz1}) into
Eq.(\ref{An}). Let us choose, for instance, $Q=0.9998\,\om_0/c$: This choice
allows the maximum value $N=158$ of $n$, as one can infer from
Eq.(\ref{N}).  Let us specify that, in such a case, one is not obliged to
use just $N=158$, but one can adopt for $N$ any values {\em
smaller} than it; more in general, any value smaller than that calculated
via Eq.(\ref{N}). Of course, on using the maximum value allowed for $N$,
one will get a better result.

\h In the present case, let us adopt the value $N=20$. In Fig.1(a) we
compare the intensity of the desired longitudinal function $F(z)$ with
that of the Frozen Wave (FW), \
$\Psi(\rho=0,z,t)$, \ obtained from Eq.(\ref{soma2}) by using the mentioned
value $N=20$.

\begin{figure}[!h]
\begin{center}
 \scalebox{1.15}{\includegraphics{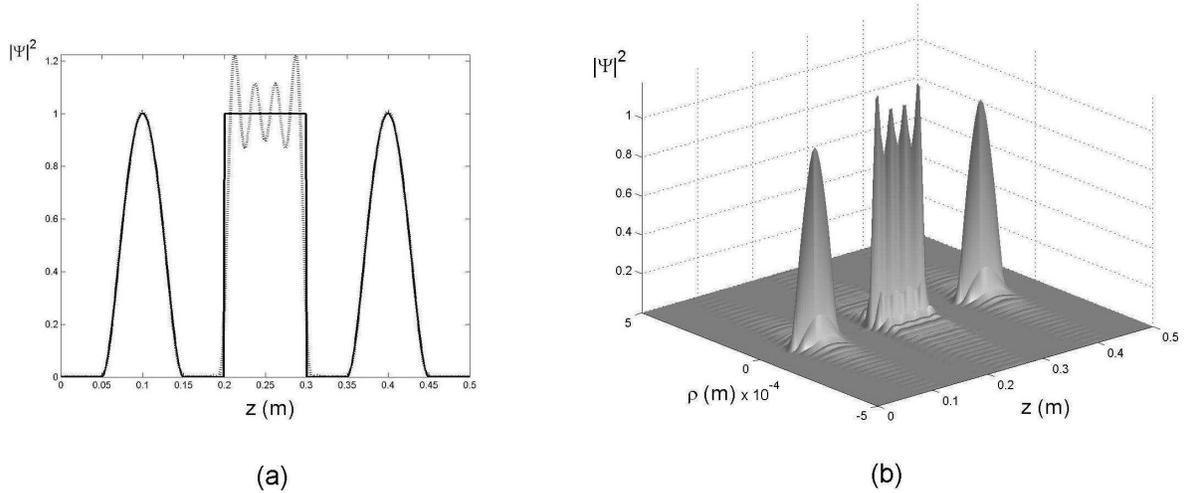}}
\end{center}
\caption{\textbf{(a)} Comparison between the intensity of the
desired longitudinal function $F(z)$ and that of our Frozen Wave (FW), \
$\Psi(\rho=0,z,t)$, \ obtained from Eq.(\ref{soma2}). The solid
line represents the function $F(z)$, and the dotted one our FW.
\textbf{(b)} 3D-plot of the field intensity of the FW chosen in
this case by us.} \label{fig2}.
\end{figure}

\h One can verify that a good agreement between the desired longitudinal
behaviour and our approximate Frozen Wave is already obtained with
$N=20$. Obviously, the use of higher values for $N$ will improve the
approximation.

\h Fig.1(b) shows the 3D-intensity of our FW, given by
Eq.(\ref{soma3}). One can observe that this field possesses the desired
longitudinal pattern, while being endowed with a good transverse localization.


\h We can expect that, for a desired longitudinal pattern of the field
intensity, on choosing smaller values of the parameter $Q$ one will get
FWs with higher {\em transverse} width (for the same number of terms
in the series (\ref{soma3})), because of the fact that the Bessel beams in
(\ref{soma3}) will possess a larger transverse wave number, and consequently
higher transverse concentrations. We can verify this expectation on
considering, for instance, a desired longitudinal
pattern, in the range $0 \leq z \leq L$, given by the function

\

 \bb
 F(z) \ug \left\{\begin{array}{clr}
 -4\,\,\dis{\frac{(z-l_1)(z-l_2)}{(l_2 - l_1)^2}} \;\;\; & {\rm in}\;\;\; l_1 \leq z \leq l_2  \\

 \\
 \;\;\;\;\;\;\;\; 0  & \mbox{in the otherwise}
\end{array} \right. \; , \label{Fz2}
 \ee

\

with $l1=L/2-\Delta L$ and $l2=L/2+\Delta L$.  Such a function has a
parabolic shape, with the peak centered at $L/2$ and a width of $2
\Delta L$.  By adopting $\lambda_0 = 0.632\,\mu$m (that is, $\om_0 = 2.98
\, 10^{15}\,$Hz), let us use the superposition (\ref{soma3})
with two different values of $Q$: we shall obtain two different
FWs that, in spite of having the same
longitudinal intensity pattern, will have different transverse
localizations.  Namely, let us consider $L=0.5\,$m and $\Delta L = L/50$,
and the two values $Q=0.99996 \om_0/c$ and $Q=0.99980 \om_0/c$.
In both cases the coefficients $A_n$ will be the same, calculated from
Eq.(\ref{An}), on using this time the value $N=30$ in the
superposition (\ref{soma3}). The results are shown in
Figures (2a) and (2b). One can observe that both
FWs have the (same) longitudinal intensity pattern, but the one
with the smaller $Q$ is endowed with the higher transverse localization.

\begin{figure}[!h]
\begin{center}
 \scalebox{1.1}{\includegraphics{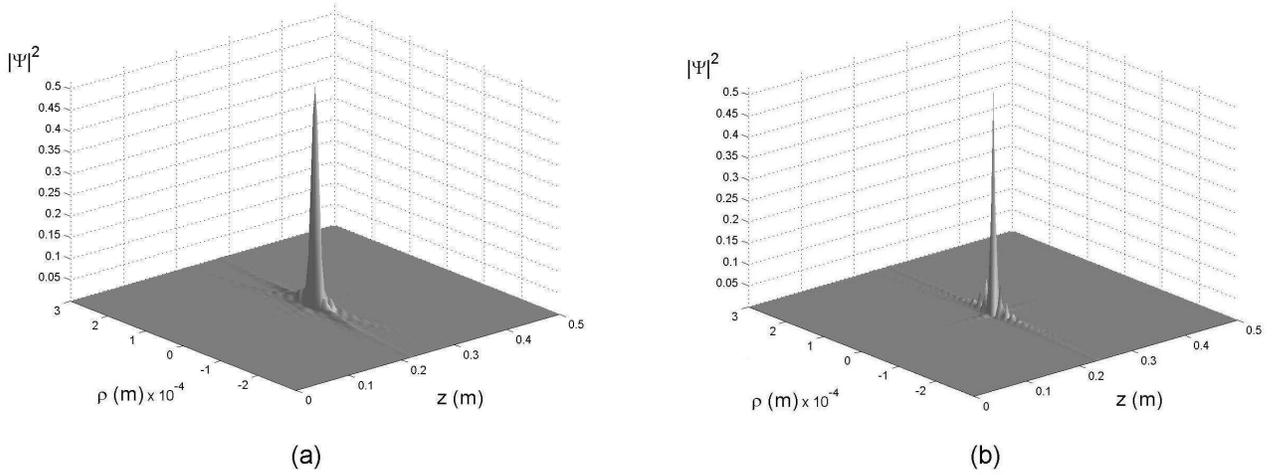}}
\end{center}
\caption{\textbf{(a)} The Frozen Wave with $Q=0.99996\om_0/c$ and
$N=30$, approximately reproducing the chosen longitudinal pattern
represented by Eq.(\ref{Fz2}). \ \textbf{(b)} A different Frozen wave,
now with $Q=0.99980 \om_0/c$ (but still with $N=30$) forwarding the same
longitudinal pattern. We can observe that in this case (with a lower
value for $Q$) a higher transverse localization is obtained.} \label{fig2}.
\end{figure}

\

{\bf 4. -- Generation of Frozen Waves}

\h Concerning the generation of Frozen Waves, we have to recall
that the superpositions (\ref{soma3}), which define them, consists of
sums of Bessel beams. Let us also recall that a Bessel beam,
when generated by finite apertures (as it must be, in any real
situations), maintains its nondiffracting properties till a
certain distance only (its field depth), given by

\bb Z \ug  \frac{R}{\tan\theta} \; , \label{ldif}\ee

where $R$ is the aperture radius and $\theta$ is the so-called
axicon angle, related with the longitudinal wave number by the
known expression[1] $\cos\theta=c\beta/\om$.


\h So, given an apparatus whatsoever capable of generating a
single (truncated) Bessel beam, we can use {\em an array} of such
apparatuses to generate a sum of them, with the appropriate
longitudinal wave numbers and amplitudes/phases (as required by
Eq.(\ref{soma3})), thus producing the desired FW. \ Here, it is
worthwhile to notice that we shall be able to generate the desired
FW in the the range $0\leq z \leq L$ if all Bessel beams entering
the superposition (\ref{soma3}) are able to reach the distance $L$
resisting the diffraction effects. We can guarantee this if $L
\leq Z_{\rm min}$, where $Z_{\rm min}$ is the field depth of the
Bessel beam with the smallest longitudinal wave number
$\be_{n=-N}=Q-2\pi N/L$, that is, with the shortest depth of
field. In such a way, once we have the values of $L$, $\om_0$,
$Q$, $N$, from Eq.(\ref{ldif}) and the above considerations it
results that the radius $R$ of the finite aperture has to be

\bb R \geq L\dis{\sqrt{\frac{\om_0^2}{c^2\be_{n=-N}^2}-1}} \ee

\h The simplest apparatus capable of generating a Bessel beam is
that adopted by Durnin et al.[6], which consists in an annular
slit located
at the focus of a convergent lens and illuminated by a
cw laser.  Then, an array of such annular rings, with the
appropriate radii and transfer functions able to yield both
the correct longitudinal wave numbers\footnote{Once a value for $Q$ has
been chosen.} and the coefficients $A_n$ of the fundamental
superposition (\ref{soma3}), can generate the
desired FW. This questions will be analyze in more detail elsewhere.

\h Obviously, other powerful tools, like the computer generated
holograms (ROACH's approach, for instance), may be used to
generated our FWs.

\

{\bf 5. -- Conclusions}

\h In this work we have shown how Bessel beams can be used to obtain
{\em stationary} localized wave fields, with high transverse localization,
whose longitudinal intensity pattern can assume any desired
shape within a chosen space interval $0\leq z \leq L$. The produced
envelope remains static, i.e., with velocity $V=0$, and because of this we
have called Frozen Waves such news solutions.

\h The present results can find applications in many
fields:$^{**}$ For instance, in the optical tweezers modelling,
since we can construct stationary optical fields with a great
variety of shapes, capable, e.g., of trapping particles or tiny
objects at different locations. This topic is being studied and
will be reported elsewhere.

\

{\bf Acknowledgements}\\

The author is very grateful to Erasmo Recami, Hugo E. H. Figueroa,
Marco Mattiuzi, C. Dartora and V. Abate for continuous discussions
and
collaboration. This work was supported by FAPESP (Brazil). \\

\

{\bf References:}\hfill\break

[1] For a review, see: E.Recami, M.Zamboni-Rached, K.Z.N\'obrega,
C.A.Dartora, and H.E.Hern\'andez-Figueroa, ``On the localized
superluminal solutions to the Maxwell equations,'' IEEE Journal of
Selected Topics in Quantum Electronics {\bf 9}, 59-73 (2003); and
references therein.

\

[2] Z. Bouchal and J. Wagner, ``Self-reconstruction effect in free
propagation wavefield,'' Optics Communications {\bf 176}, 299-307
(2000).

\

[3] Z. Bouchal, ``Controlled spatial shaping of nondiffracting
patterns and arrays,'' Optics Letters {\bf 27}, 1376-1378 (2002).

\

[4]J. Rosen and A. Yariv, ``Synthesis of an arbitrary axial field
profile by computer-generated holograms,'' Optics Letters {\bf
19}, 843-845 (1994).

\

[5] R. Piestun, B. Spektor and J. Shamir, ``Unconventional light
distributions in three-dimensional domains,'' Journal of Modern
Optics {\bf 43}, 1495-1507 (1996).

\

[6] J. Durnin, J. J. Miceli and J. H. Eberly, ``Diffraction-free
beams,'' Physical Review Letters {\bf 58}, 1499-1501 (1987).

\end{document}